\documentclass[12pt]{iopart}
\usepackage{iopams}  
\usepackage{graphicx}
\usepackage{xspace}
\usepackage{booktabs}
\usepackage{cite}
\usepackage{lscape}

\newcommand{\lisix}{\textsuperscript{6}Li\xspace}

\usepackage{graphicx}

\begin{document}

\title[Probing the cluster structure of $^6$Li with the nuclear reaction $^6$Li + $^{12}$C at 68 MeV]{
Probing the cluster structure of $^6$Li with the nuclear reaction $^6$Li + $^{12}$C at 68 MeV
}

\author{Urazbekov~B~A$^{1,2}$, 
Almanbetova~E~K$^{3}$
Azhibekov~A$^{3,4}$,
Baimurzinova~B~S$^5$
Dyussebayeva~K$^{6}$
Issatayev~T$^{2,3,*}$,
Janseitov~D~M$^{2}$, 
Lukyanov~S~M$^{3}$, 
Penionzhkevich~Yu~E$^{3}$, 
Mendibayev~K$^{2,3}$,
Zholdybayev~T~K$^{2,6}$}

\address{$^1$Gumilyov  Eurasian National University,  2 Satpayev Str.,Astana 010000, Kazakhstan}
\address{$^2$Institute of Nuclear Physics, 1 Ibragimov Str.,  Almaty 050032, Kazakhstan}
\address{$^3$Flerov Laboratory of Nuclear Reactions, JINR, 20 Joliot Curie Str.,  Dubna 141980, Russia} 
\address{$^4$Korkyt-ata State University,  29A Aiteke-Byi Str,  Kyzylorda 120000, Kazakhstan}
\address{$^5$SDU University, 1/1 Abylaikhan Str., Kaskelen,  Almaty 040900, Kazakhstan}
\address{$^6$Al-Farabi Kazakh National University, 71 Al-Farabi Ave., Almaty 050040, Kazakhstan}
\ead{$^*$issatayev@jinr.ru}
\vspace{10pt}
\begin{indented}
\item[July 2025]
\end{indented}

\begin{abstract}
This work presents a combined experimental and theoretical investigation of the nuclear reaction $^6$Li + $^{12}$C at a laboratory energy of 68 MeV. 
The reaction products are identified via the standard $\Delta E$–$E$ technique. Angular distributions are constructed for the elastic, inelastic, and deuteron transfer channels by detecting emitted particles -- $^6$Li and $\alpha$.

Elastic and inelastic scattering of $^6$Li of $^{12}$C are analyzed using both the Optical Model and Coupled channels approaches, with the interaction described by a double-folding potential. This potential is calculated based on the three-body wave function of $^6$Li. 
Pronounced coupled-channel effects are observed, which modify the potential and allow accurate reproduction of the experimental cross sections. The resulting polarized potentials provide a more precise description of the initial-state interaction for further reaction modelling.

The deuteron transfer channel, $^{12}$C($^6$Li, $\alpha$)$^{14}$N, is studied using the Coupled Reaction Channels method. The coupling between the transfer and elastic channels is implemented using the three-body wave function of $^6$Li. As an alternative, a regular wave function constructed with a phenomenological Woods–Saxon potential is also employed.
Comparison between the calculated differential cross sections and experimental data reveals a more complex and nuanced reaction mechanism, which supports the cluster structure of $^6$Li.

\end{abstract}

%
%
%
%
%
\clearpage

\section*{Introduction}
With the development of detection systems, increased computing power, and the emergence of new theoretical approaches, the study of light nuclei remains a central topic in nuclear structure research. 

In particular, the nucleus $^6$Li is a light nucleus characterized by a well-established cluster structure, typically modelled as a bound system of an alpha particle and a deuteron ($\alpha + d$). 
This configuration plays a crucial role in the internal dynamics of the nucleus and distinguishes $^6$Li from other light systems \cite{NSR2022KI18, NSR2021AM06, NSR2020RO08}. 
Due to its relatively weak binding energy and extended spatial correlations between constituents, a full three-body description is often necessary to capture its physical properties, as simpler two-body models may neglect essential degrees of freedom. 
As such, $^6$Li represents a valuable system for exploring clustering phenomena and few-body interactions in nuclear physics. 
Recent studies, both theoretical and experimental, provide further evidence of these unique features \cite{NSR2021EG01, NSR2024UR01, NSR2023SO11, NSR2023PE15}.

The clustered nature of $^6$Li has a significant impact on its behavior in nuclear reactions, including elastic and inelastic scattering and cluster transfer processes. 
Experimental studies involving $^6$Li beams have revealed features that require theoretical approaches to explicitly account for cluster dynamics. 
In particular, the use of realistic three-body wave functions has been shown to improve agreement with reaction observables \cite{zhukov1993bound, braaten2006universality, jensen2004structure}. 
Thus, investigations of $^6$Li reactions provide key insights into the interplay between nuclear structure and reaction mechanisms, and serve as a testing ground for modern few-body and cluster-based theoretical models \cite{lin1995hyperspherical,kanada1995structure, kukulin1984detailed, bang1979realistic}.
In particular, in Ref. \cite{zhukov1993bound}, tests were performed on the $^6$Li nucleus to validate the three‑body model and the adopted interactions.
The effects of the Pauli principle, various $NN$ and $\alpha$–$N$ potentials, as well as the convergence and structure of the wave functions, were taken into account. 
Binding energies, radii, the component composition of the wave functions in the $LS$ coupling scheme, and the hypermomentum decomposition were calculated. 

The internal structure of light nuclei also plays an important role in shaping nuclear reaction mechanisms \cite{starastsin2021structures, azhibekov2024study, nauruzbayev2017structure,lukyanov2016cluster, umbelino2019two}, especially when one of the colliding nuclei possesses a well‑established cluster structure.
In the work \cite{oganessian1999dynamics}, a four‑body model (target nucleus + $\alpha$‑core + two valence neutrons) was developed and applied to describe two‑neutron transfer reactions involving Borromean nuclei, primarily $^{6}$He, and based on this, data on $^{6}$He–$^{4}$He collisions were analyzed.
A realistic three‑body bound‑state wave function of $^{6}$He, obtained through expansion over hyperspherical harmonics with inclusion of \textit{n–n} and \textit{n–}$\alpha$ interactions, was used in the calculations.
The results showed that the cross sections strongly depend on the internal structure of $^{6}$He: the calculated cross sections indicate that the “dineutron” component—where two neutrons are close to each other outside the $\alpha$‑core -- gives the dominant contribution to the transfer, whereas the “cigarlike” configuration -- where the two neutrons are much farther apart than the $\alpha$‑core -- is noticeable only at larger angles.

Motivated by these findings, we aim to explore the cluster structure of $^6$Li in the nuclear  reaction $^6$Li + $^{12}$C  at the laboratory  energy of 68 MeV.
In particular,  our interest focuses on the deuteron‑transfer channels originating from the $^6$Li nucleus. 
Based on the well‑established cluster structure, we employ a three‑body model of $^6$Li. This work focuses on two central questions: in what way the cluster transfer contributes to the nuclear reaction, and how the overall reaction mechanism is affected when the three‑body structure is not taken into account.

This paper is a continuation of the series of works \cite{Azhibekov2024, urazbekov2019clusterization, urazbekov2016manifestation, urazbekov2021application, urazbekov2024reactions} devoted to the study of cluster effects in nuclear reactions. 
In this context, the present work extends previous analyses by focusing on the $^6$Li + $^{12}$C  reaction at 68 MeV, with particular attention to the deuteron‑transfer channel. 

The paper is organized as follows.
In the first section, we describe the experimental procedure, the measurement conditions, and the methods used to obtain and process the data.
The second section introduces the three-body wave function employed in this work and outlines the procedure used to derive the nuclear matter density distribution of $^6$Li, which serves as input for the subsequent reaction calculations.
The third section is devoted to interpreting the calculated results for elastic and inelastic scattering in comparison with the experimental cross sections.
The fourth section focuses on the analysis of the experimental data for the deuteron‑transfer reaction and its correlation with the three‑body structure of $^6$Li.
Finally, the main conclusions and outlook are summarized in the last section.

\begin{figure}
    \centering
    \includegraphics[scale=0.1]{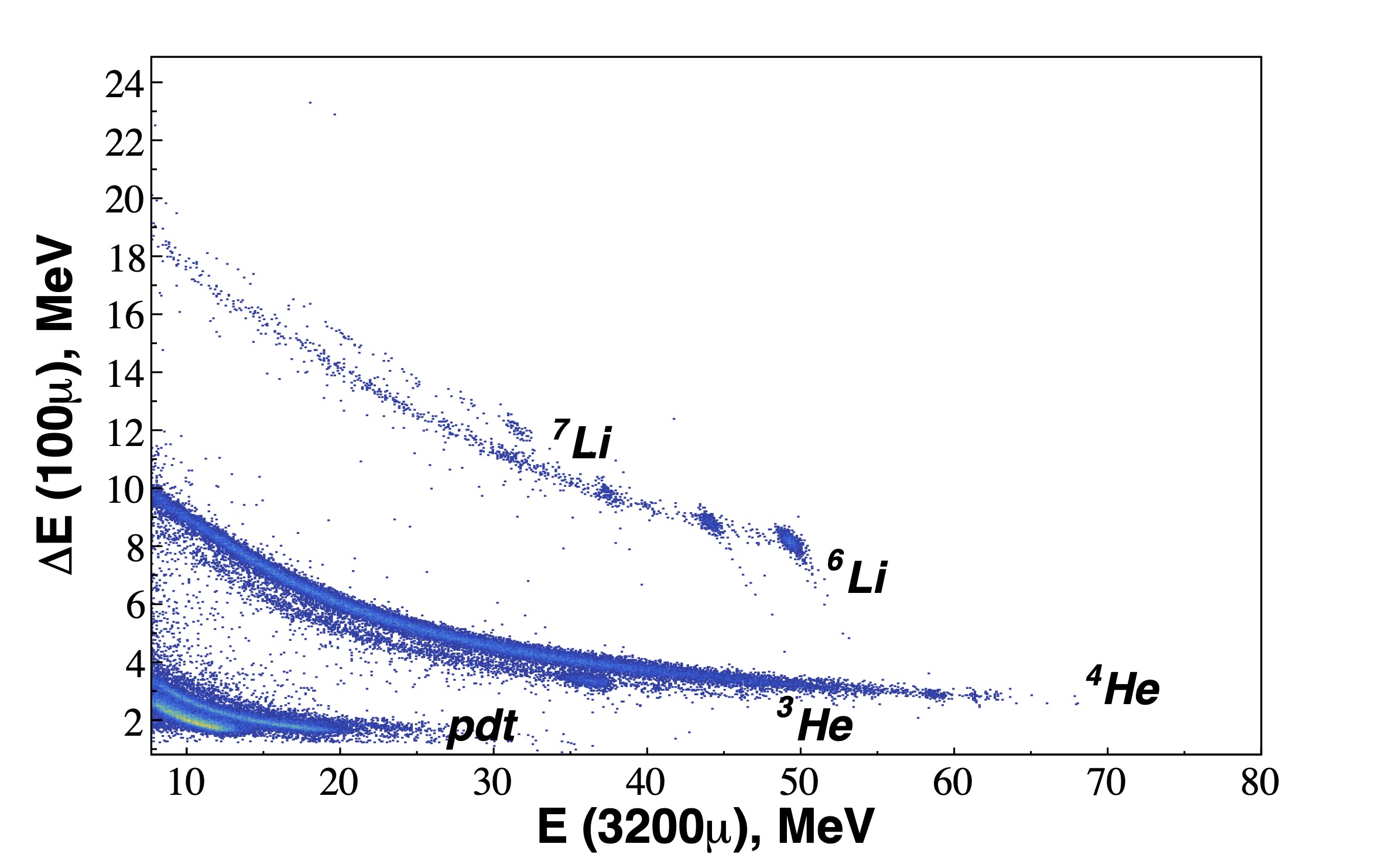}
\caption{Typical two-dimensional particle identification spectrum ($\Delta E$–$E$) for the reaction products of the $^6$Li + $^{12}$C reaction, measured at a laboratory angle of $\theta_{\textrm{lab}} = 26^\circ$. The data were obtained using one of the silicon telescopes. The loci corresponding to different isotopes, including $^4$He, $^6$Li, and $^7$Li, are clearly separated.}
    \label{fig:spectrum}
\end{figure}

\begin{figure}
    \centering
    \includegraphics[scale=0.1]{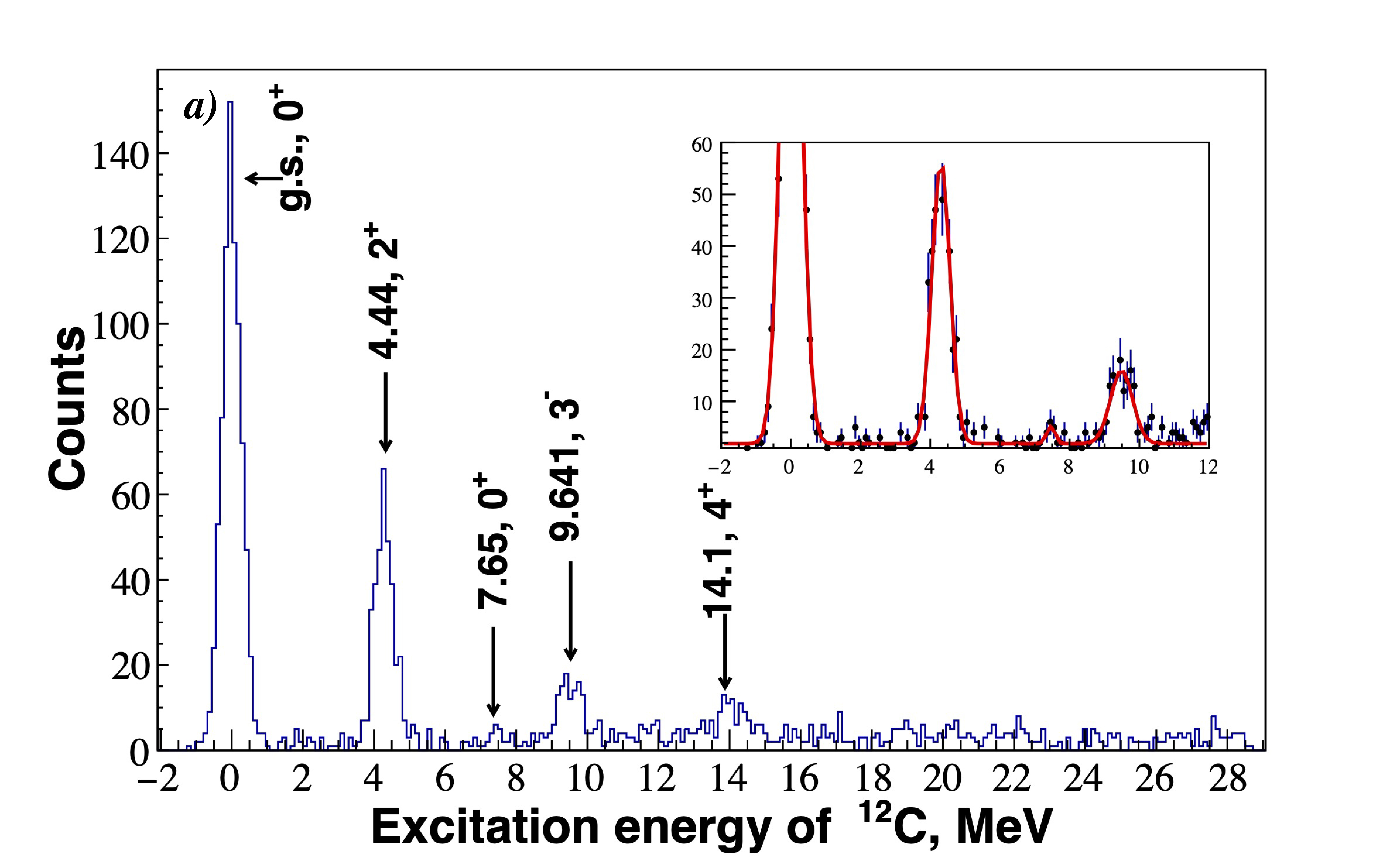}
        \includegraphics[scale=0.1]{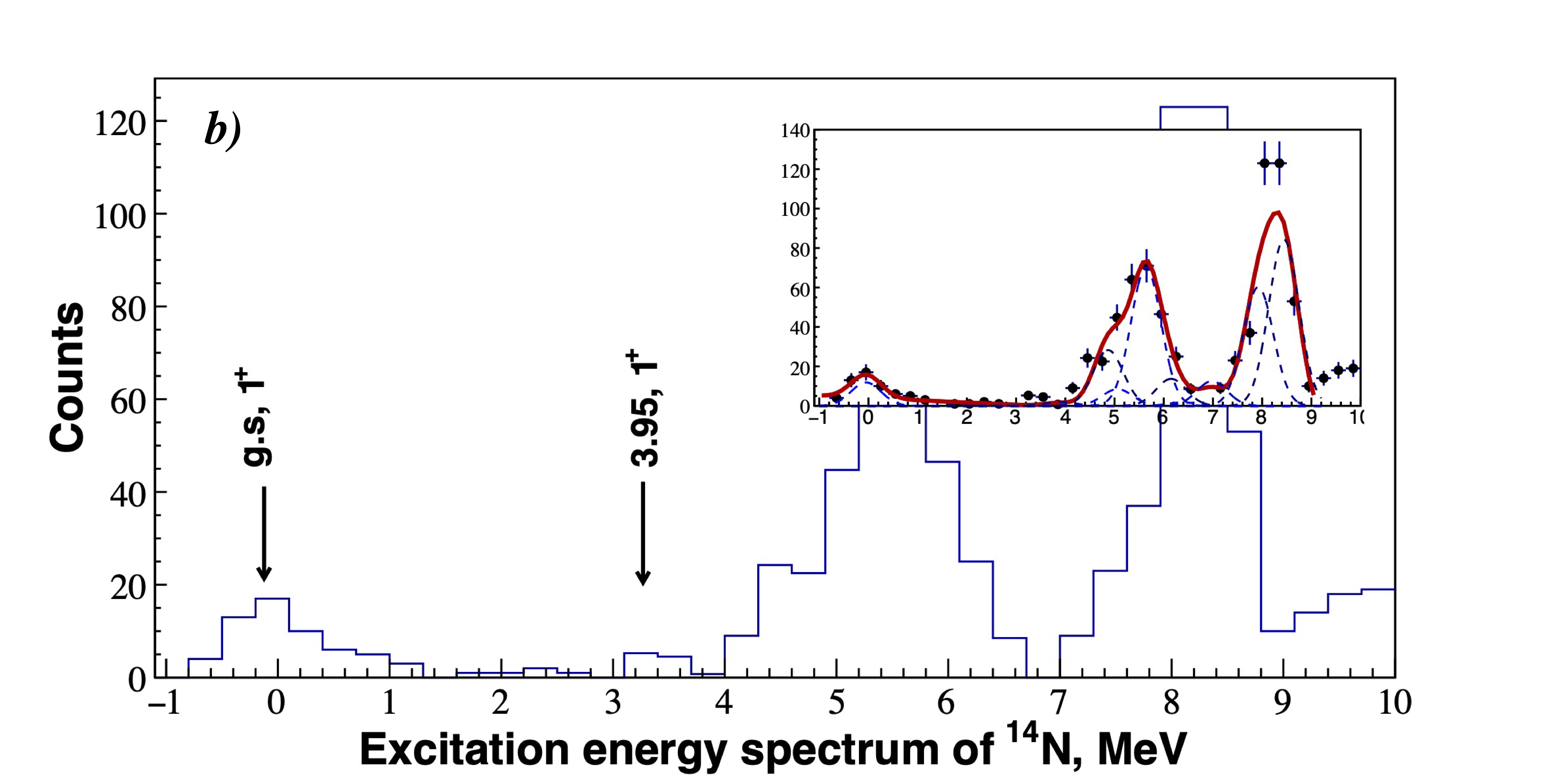}
\caption{\textit{a)} Excitation energy spectrum of $^{12}$C obtained from the $^6$Li + $^{12}$C reaction at $E_{\textrm{lab}} = 68$~MeV. The spectrum displays peaks corresponding to elastic scattering (g.s.\ $0^+$) and inelastic scattering to excited states at 4.44~MeV ($2^+$), 7.65~MeV ($0^+$), 9.641~MeV ($3^-$), and 14.1~MeV ($4^+$). The inset illustrates the peak fitting procedure used to extract the yields for each state.
\textit{b)}  Excitation energy spectrum for the $^{12}$C($^6$Li,~$\alpha$)$^{14}$N reaction channel at $E_{\textrm{lab}} = 68$~MeV. The ground state (g.s.\ $1^+$), which is the focus of the theoretical analysis in this work, is clearly identified. The spectrum also reveals the population of the 3.95~MeV state and several higher-lying states. The inset shows a multi-peak fit used to extract the contributions from individual excited states.
}

    \label{fig:elasticAndTransferSpectra}
\end{figure}

\section{Experimental procedure}

Experiments with a $^6$Li beam on a $^{12}$C target were conducted at the Flerov Laboratory of Nuclear Reactions, Joint Institute for Nuclear Research, Dubna, Russia. A $^6$Li beam with an energy of 68~MeV was accelerated by the U-400 cyclotron and directed to the reaction chamber of the high-resolution MAVR spectrometer. The magnetic optics of the U-400, combined with a system of diaphragms, formed the beam profile. A profilometer installed upstream of the reaction chamber confirmed a beam spot of approximately $5 \times 5$~mm$^2$ on the target, with a typical current of 30~nA. The beam charge was monitored using a Faraday cup and cross-checked via elastic scattering measurements.

The beam was incident on a self-supporting $^{12}$C foil with a thickness of 5~$\mu$m and a purity greater than 99\%. No contaminant isotopes were observed in the energy spectra.

Particle identification was carried out using the $\Delta E$--$E$ technique, measuring energy loss in the first two detector layers and the residual energy in the third. Four three-layer silicon telescopes were employed with thicknesses of $50~\mu$m--$100~\mu$m--$3200~\mu$m for forward angles and $12~\mu$m--$700~\mu$m--$3200~\mu$m for backward angles. The solid angle and angular acceptance were approximately 0.03~msr and $0.35^\circ$ for forward-angle telescopes, and 0.2~msr and $0.9^\circ$ for backward-angle telescopes. This configuration enabled clear identification of reaction products over a wide energy range, from helium to boron isotopes.

Examples of particle identification matrices obtained with one of the telescopes are shown in Fig.~\ref{fig:spectrum}, demonstrating excellent separation of all detected species. By selecting appropriate gates on the particle loci in the identification matrix (Fig.~1), energy spectra for various reaction channels were extracted. Figure~\ref{fig:elasticAndTransferSpectra}~(\textit{a}) presents the energy spectrum for outgoing $^6$Li particles, showing peaks corresponding to elastic scattering from the $^{12}$C ground state and inelastic scattering to its excited states. Similarly, Fig.~\ref{fig:elasticAndTransferSpectra}~(\textit{b}) shows the excitation energy spectrum for the $^{12}$C($^6$Li,~$\alpha$)$^{14}$N transfer reaction, where the ejectiles are $\alpha$ particles. The spectrum reveals several populated states in the residual $^{14}$N nucleus, though the present analysis focuses only on the ground-state transition.

The energy spectra and angular distributions were measured using an inclusive method. The overall energy resolution was primarily influenced by the intrinsic energy spread of the $^6$Li beam and uncertainties in energy loss within the $^{12}$C target. The resulting energy resolution was approximately 500~keV for ejectiles with $Z = 1$--3, and around 1~MeV for those with $Z = 4$--6.

\begin{figure}
    \centering
    \includegraphics[scale=0.4]{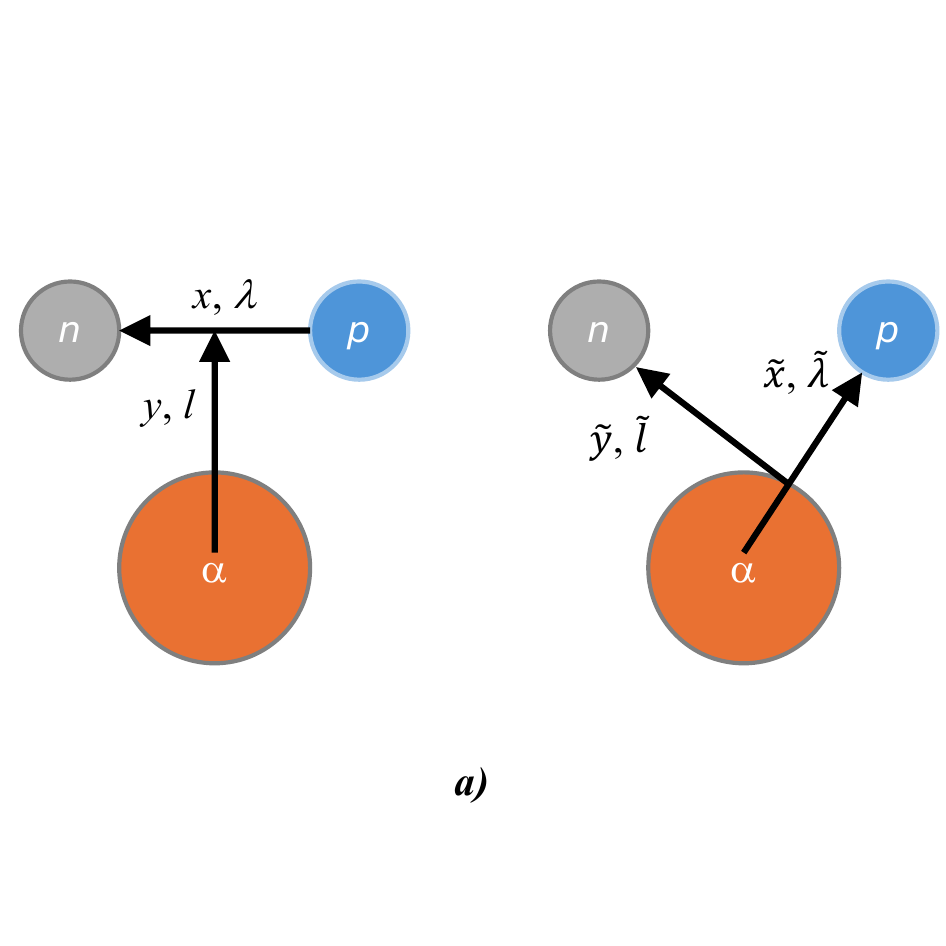}~
        \includegraphics[scale=0.7]{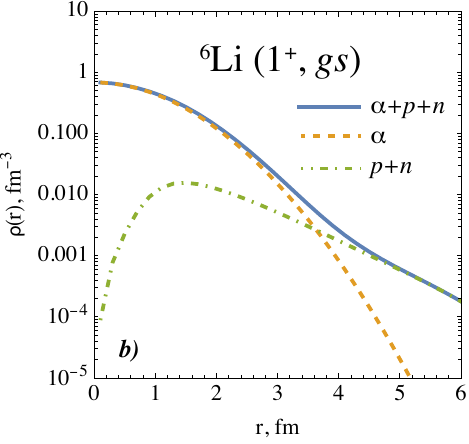}
        \caption{
        \textit{a)} Schematic representation of the three-body model of $^6$Li with relative Jacobi coordinates $x$ and $y$ for the system of an $\alpha$ particle and two nucleons. The "\textit{T}"-type configuration is shown on the left, and the "\textit{Y}"-type on the right.
\textit{b)}
Nuclear matter density distribution functions $\rho(r)$ for $^6$Li calculated within the three-body model: solid line -- total density, dashed line -- $\alpha$ cluster, dot‑dashed line -- \textit{p+n} cluster.        
        }
    \label{fig:3bJacobi}
\end{figure}

\section{The three body model}
In this work, the $^6$Li nucleus is modelled as a three-body system consisting of one $\alpha$ particle, proton, and neutron.
The quantum state of this system is described by the wave function $\Psi^{JM}(x,y)$, where $J$ is the total angular momentum and $M$ is its projection. The wave function is constructed using the Jacobi relative coordinates $x$ and $y$, which eliminate the center-of-mass motion.
The coordinate configuration is illustrated in Fig.~\ref{fig:3bJacobi}~(\textit{a}).
The three-body wave function is constructed based on pairwise interactions in the $\alpha$–$N$ and $N$–$N$ subsystems. Specifically, the $\alpha$–$N$ interaction is modeled using a parity-dependent potential fitted to reproduce scattering data \cite{kukulin1995detailed}. This potential accurately reproduces phase shifts for proton–$\alpha$ scattering up to 20 MeV. The $N$–$N$ interaction is described using the Reid soft-core potential \cite{day1981three}.
It is worth noting that the $\alpha$–$N$ system contains a forbidden state in the $S$-wave channel. This issue is addressed using the orthogonalising pseudo-potentials technique \cite{kukulin1978orthogonal}, which projects out the forbidden state during the variational calculation.

The wave function $\Psi^{JM}(x,y)$ used as adopted from Ref. \cite{kukulin1984detailed} provides a good description of $^6$Li, especially its low-lying energy spectrum and electromagnetic properties, which makes it a physically well-grounded and reliable basis for further nuclear reaction calculations.

The nuclear matter density distribution of $^6$Li in the three-body model is written as \cite{UrazbekovThesis}:
\begin{equation}
\rho(r) = \sum_i \rho_i(r),
\label{rho3b}
\end{equation}
where the sum is taken over all the clusters in the system. For each cluster, the density is calculated as :
\begin{equation}
\rho_i(r) = \langle \Psi^{JM}(x,y) | ~\hat{\rho}_i ~| \Psi^{JM}(x,y) \rangle,
\end{equation}
where $\hat{\rho}_i$ is the density operator defined as:
\begin{equation}
\hat{\rho}_i = \cases{
\delta(y-y_0 r), \textrm{ \textit{i} nucleon} \\
\rho_\alpha(y_i-y_0 r), \textrm{for \textit{i}}~\alpha~\textrm{particle}.
}
\end{equation}
where $y_0$ takes the value $5/6$ for nucleons and $1/3$ for the $\alpha$-cluster. 
In this work, nucleons are treated as point-like, whereas the $\alpha$ particle is assumed to have an internal structure described by a Gaussian distribution that reproduces the root-mean-square (rms) radius of 1.461 fm \cite{satchler1979folding}.
To facilitate the treatment of nucleons, the Jacobi coordinate system is transformed from the “\textit{T}”-type to the “\textit{Y}”-type (see Fig.~\ref{fig:3bJacobi}~(\textit{a})).

The nuclear matter density distribution functions were calculated with the three body wave function parameters taken from Ref. \cite{pomeranPC} and are shown in Fig.~\ref{fig:3bJacobi}~(\textit{b})  in terms of clusters contributions. 
In this work, we employed the $S$‑wave component of the three‑body configuration set, as it accounts for the dominant contribution, about 89\% of the total weight.
A distinctive feature of the obtained results is the extended tail of the density distribution. This behavior is due to the properties of the valence nucleons in the three-body system. In particular, the density of the $\alpha$ cluster decreases rapidly with increasing radius $r$, while the density of the valence nucleons falls off more slowly. This indicates a stronger contribution from the $p+n$-cluster component.
Another notable feature is a local maximum in the nucleon density at $r \approx 1.5$ fm, which reflects the fact that the valence nucleons are spatially separated from the center of mass.

Using the obtained nuclear matter density distribution for $^6$Li, we can estimate its rms radius. The calculated rms radius is 2.49 fm,  which falls within the range of values 2.54 $\pm$ 0.03 reported in  Ref. \cite{tanihata1985measurements}.

\begin{figure}
    \centering
    \includegraphics[scale=0.8]{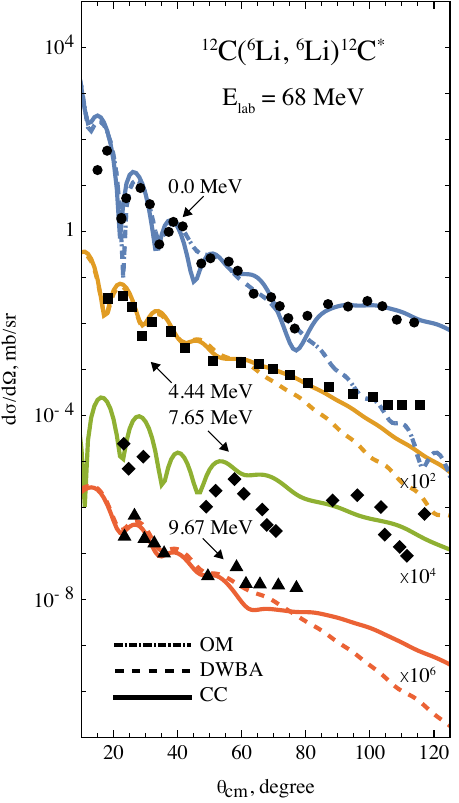}
    \caption{Experimental angular distributions of elastic (circles) and inelastic scattering at excitation energies of 4.44 MeV (squares), 7.65 MeV (diamonds), and 9.67 MeV (triangles) resulting from the $^6$Li + $^{12}$C$^*$ reaction at $E_{lab} = 68$ MeV, shown in comparison with different model calculations: solid line -- CC method, dashed -- DWBA, dash-dotted -- OM.}
    \label{fig:elAndInel}
\end{figure}

\section{Elastic and inelastic channels}
\subsection*{Optical Model}
The differential cross sections of the elastic scattering of \lisix from the nucleus $^{12}$C at the laboratory energy of 68 MeV are treated within the Optical Model (OM) framework. 
The optical potential used in the OM calculations was taken in the form:

\begin{equation}\label{eqn:OP}
\begin{array}{l}
 U(R)=-V(R)-iW(R)+V^C(R),
\end{array}
\end{equation}
where $R$ is the distance between the \lisix  and $^{12}$C, and $V, W$ are the real and imaginary volume potential terms and $V^C$ is the Coulomb potential.

The real volume potential $V(R)$ can be obtained within the double-folding model~\cite{satchler1979folding}:
\begin{equation}
V(R) \equiv N_R V^{DF}({R}) = \int \int d\textbf{r}_p, d\textbf{r}_t, \rho_p(\textbf{r}_p), V_{NN}(\textbf{R} - \textbf{r}_p + \textbf{r}_t), \rho_t(\textbf{r}_t),
\end{equation}
where $\textbf{r}_p$ and $\textbf{r}_t$ are internal coordinates of the projectile and target nuclei, respectively; $\rho_p$ and $\rho_t$ are their nuclear matter density distributions; and $V_{NN}$ is the effective nucleon–nucleon interaction; $N_R$ is a free parameter adjusted to reproduce the observed reaction dynamics.
In this work, the projectile density $\rho_p$ is obtained from the three-body wave function of $^6$Li (Eq. \ref{rho3b}), whereas the target density $\rho_t$ is modeled with a parametrized Fermi distribution for $^{12}$C, constrained by the experimental rms charge radius of 2.47 fm (see, e.g., Ref. \cite{12Ccharge}).

The imaginary part of the optical potential can be represented by a Woods–Saxon form \cite{woods1954diffuse}:
\begin{equation}
W\left( R \right) = 
\frac{W_0}{1+ \exp \left( \frac{R-R_W}{a_W} \right) },
\label{eq:imPot}
\end{equation}
where $W_0$ is the depth of the potential, $R_W$ is the radius parameter, and $a_W$ is the diffuseness parameter, while the Coulomb interaction is taken as the interaction of a point-charge with a uniformly charged sphere
\begin{eqnarray}
\label{coul}
V^C(R)=
\cases{
\frac{Z_1 Z_2 e^2}{2 R_C} \left( 3- \frac{R^2}{R_C^2} \right), & for  R $\leq R_C$, \\
\frac{Z_1 Z_2 e^2}{R}, & for  R $> R_C$ .
}
\end{eqnarray}

\begin{figure}
    \centering
    \includegraphics[scale=0.7]{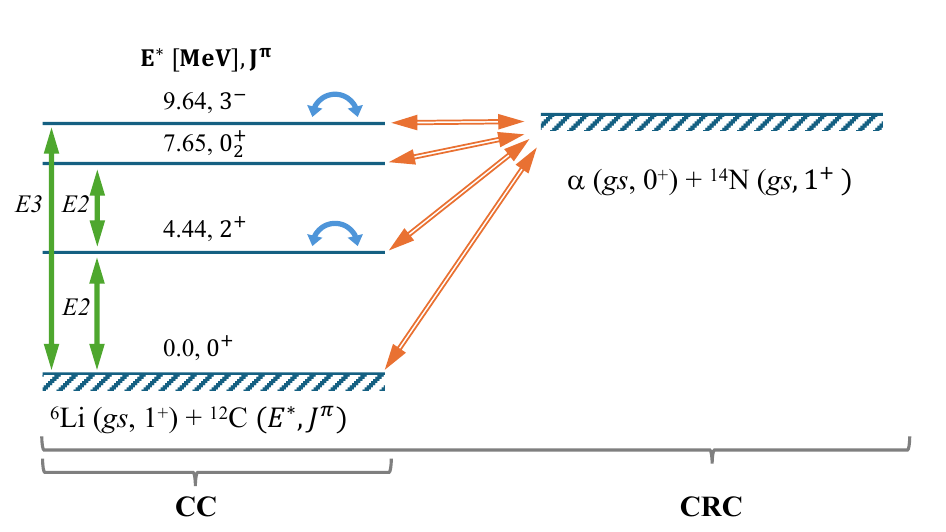}
    \caption{Coupling scheme used in the CC calculations for $^{12}$C($^6$Li,~$^6$Li')$^{12}$C$^*$ inelastic scattering and in the CRC calculations for $^{12}$C($^6$Li,~$\alpha$)$^{14}$N.
    Double‑headed arrows indicate $E\lambda$ transitions, backward‑pointing arrows indicate spin-reorientation, and double-lined arrows indicate reaction transitions.}
    \label{fig:couplScheme}
\end{figure}

The double‑folding (DF) potential is obtained using the Python‑based code \texttt{BiFOLD}, which implements the nuclear density functions mentioned above and the nucleon–nucleon interaction of the CDM3Y‑6 type reported in Ref. \cite{khoa2001alpha}. 
The resulting DF potential is then used to analyze the elastic scattering channel within the OM using the \texttt{FRESCO} code \cite{thompson1988coupled}. The imaginary part of the DF potential is fitted by $\chi^2$ minimization, with the initial global parameters taken from Ref. \cite{xu2018li} and tabulated in Tab. \ref{potpar}.

The results of numerical calculations within the OM for the elastic scattering reaction $^{12}$C($^6$Li, $^6$Li)$^{12}$C at a laboratory energy of 68 MeV, in comparison with the experimental data, are shown in Fig. \ref{fig:elAndInel}. The DF potential used within the OM reproduces the experimental data well, except in the angular region beyond 90$^\circ$. The good agreement at forward angles confirms the validity of the modeled interactions between the colliding nuclei based on the three‑body structure of the $^6$Li nucleus.

\subsection*{Coupled channels}

 One specific feature of the results is the remaining discrepancy at backward angles in the elastic scattering channel.
 Indeed, when the Coupled Channels (CC) approach is employed together with the coupling scheme shown in Fig.~\ref{fig:couplScheme}, which also includes backward couplings in addition to the Distorted Wave Born Approximation (DWBA), the calculated differential cross sections reproduce the experimental elastic scattering data even at backward angles, providing a more accurate overall description. 
 Detailed analyses have shown that a strong coupled-channels effect arises primarily from the inelastic scattering channel at 4.44 MeV, which leads to an increased elastic scattering cross section at angles beyond 90$^\circ$.

Electromagnetic transition strengths for Coulomb excitation coupling were adopted from Ref. \cite{kelley2017energy}.
The deformation length for the transition in $^{12}$C from the ground state to the first excited $2^+$ state at 4.44~MeV reproduces the experimental differential cross sections well when taken in the range of $1.75 \pm 0.15$~fm. 
The obtained deformation length lies within the values reported for $^4$He projectiles \cite{yasue1983deformation}, namely 1.95~fm at a laboratory energy of 139~MeV and 1.37~fm at a laboratory energy of 65~MeV.
Particular interest is also associated with the deformation length for the octupole transition in $^{12}$C, i.e., from the ground state to the excited state at 9.64~MeV. 
The coupled–channels analysis shows that the value of $0.70 \pm 0.07$~fm provides a good description of the experimental cross sections. 
Indeed, this value is consistent with the results obtained using the deformation parameter $\beta_3 = 0.26$ \cite{aspelund1975elastic}, which generates the deformation length 
$\delta_3 = \beta_3 \cdot 1.2 A^{1/3} = 0.71$~fm.
These results were obtained within the coupled–channels method with the same projectile at laboratory energies of 60 and 77~MeV.

The CC model provides a satisfactory reproduction of the experimental data, except at angles beyond 90$^\circ$ for the channel with the 4.44 MeV excitation and beyond 60$^\circ$ for the channel with the 9.67 MeV excitation. 
This discrepancy likely arises from omitted couplings to higher-lying states in the CC coupling scheme.

The inelastic channel with the 7.65~MeV excitation also deserves special attention, as it corresponds to the well-known Hoyle state. 
The coupling does not proceed directly from the ground state due to the forbidden $E0$ transition (see Fig.~\ref{fig:couplScheme}), but rather through the intermediate excited state at 4.44~MeV. 
As a result, the definition of the deformation parameter with respect to the ground state loses its physical meaning. 
This most likely explains why the present model is insufficient to reproduce the overall angular distribution for this transition. 
The analysis clearly indicates the involvement of an additional inelastic mechanism, whose interference with the considered channels would provide a more complete description. 
For a more accurate and comprehensive treatment of this state, a more advanced modelling framework is required, one that explicitly incorporates the $3\alpha$ cluster structure.

Explicitly implemented within the CC framework, inelastic channels lead to noticeable coupled‑channels effects. These effects cause  13\% increase in the depth of the DF potential and 28\% reduction in the depth of the imaginary part of the optical potential (see Tab.\ref{potpar}). 
The modified, or polarized, potential represents how the interaction between the colliding nuclei is effectively changed by the coupling to inelastic excitation channels.
Such potentials yield a more accurate representation of the initial-state interaction and are consistent with previous findings in Refs.\cite{thompson1989threshold, urazbekov2019clusterization, harakeh1980strong}.

\begin{table*}[bp]
\footnotesize
\caption{\label{potpar} Optical potential parameters used in the OM, CC and CRC calculations.}
\center
\begin{tabular}{@{}lllllllllllll@{}}
\toprule
~& $N_R$ & $V_0$ [MeV]  & $r_V^{a)}$ [fm] & $a_V$ [fm] & $W_0$ [MeV] & $r_W^{a)}$ [fm]& $a_W$ [fm] & $r_C^{a)}$ [fm]  & $\chi^2/N$ \\ 
 \midrule
 \lisix+$^{12}$C & 0.61  & --   &  -- & -- & 23.35 & 1.528 & 0.773 & 1.67    &  7.41 \\
~ &0.69$^{b)}$ &     &   &  & 15.31$^{b)}$ & ~& ~ & ~     & ~~ ~\\
 $\alpha$+$^{14}$N & --   &187.7& 0.880 & 0.875 & 35.0 & 0.516 & 1.313 & 1.35 & 8.32\\
 ~ &    &176.6 $^{c)}$& ~ & ~ & 50.0$^{c)}$ & ~ & ~ & ~ &~~~ \\
\bottomrule
\end{tabular}\\
\scriptsize
\flushleft
$^{a)}$ $r_i = R_i A^{-1/3}_t $ \\
$^{b)}$ used in the CC and CRC \\
$^{c)}$ used in the CRC 
\end{table*}

\begin{figure}
    \centering
    \includegraphics[scale=0.6]{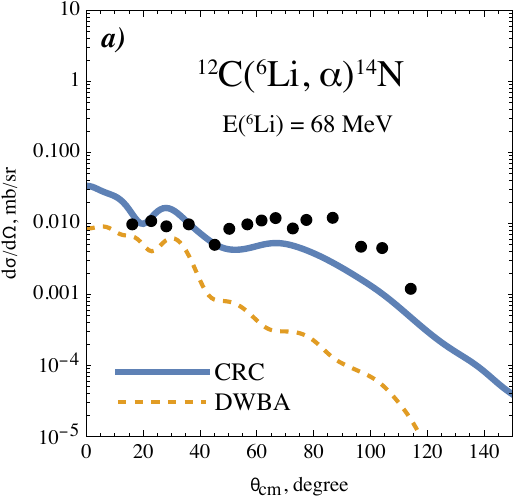} ~
        \includegraphics[scale=0.6]{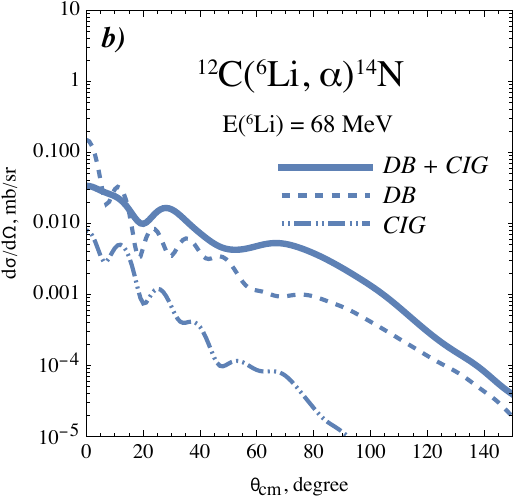}

    \caption{Experimental cross-section data for the nuclear reaction $^{12}$C($^6$Li, $\alpha$)$^{14}$N compared with various reaction models.
(\textit{a}): cross sections calculated within the CRC approach (solid line) and the DWBA approach (dashed line);
(\textit{b}): the same CRC cross sections as in the left panel, but shown in terms of contributions from two different structural configurations of $^6$Li: the DB (dashed line) and the CIG (dot-dashed line).
}
    \label{fig:deuteronransferChannel}
\end{figure}

\section{Deuteron transfer channel}
\subsection*{Coupled Reaction Channels}

The nuclear reaction $^{12}$C($^6$Li, $\alpha$)$^{14}$N is analyzed within the framework of the coupled‑reaction‑channels (CRC) approach \cite{thompson1988coupled}. 
The coupling scheme employed in the CRC framework is shown in Fig. \ref{fig:couplScheme}.
In the analysis, the initial-state interaction is chosen to be the polarised the double‑folding (DF) potential, while the final-state interaction is represented by the Woods–Saxon potential for both the real and imaginary parts. 
The parameters were adjusted to reproduce the experimental differential cross sections for $\alpha$ + $^{14}$N elastic scattering at a laboratory energy of 54~MeV~\cite{abele1987measurement}.  
The real part is expressed as follows:  
\begin{equation}
V(R) = \frac{V_0}{1 + \exp \left( \frac{R - R_V}{a_V} \right) },
\label{eq:wsReal}
\end{equation}
while the imaginary volume potential has the same shape as that used for the $^6$Li + $^{12}$C system (Eq.~\ref{eq:imPot}).
 
In the calculations, the coupling potentials were constructed using the prior form.
The potential of the final-state interaction was chosen to be the core-core potential in the remnant part, since this kind of potentials is observed to be more dependent on the projectile than the target nuclei.

\begin{table*}[tp]
\footnotesize
\caption{\label{tab:sa} 
Two-nucleon spectroscopic amplitudes, $A^{SM}$, used in the CRC calculations for the overlaps $\langle ^{14}\mathrm{N}(J_f)  | ^{12}\mathrm{C}(J_i) \rangle$. The transferred proton and neutron are assumed to occupy the shells $nlj(\rho)$ and $nlj(\eta)$, respectively, and are coupled to the total angular momentum $J$.
}
\center
\begin{tabular}{@{}lllllr|lllllr@{}}
\toprule
$J_f$  & $J_i$   &  $nlj(\rho)$& $nlj(\eta)$ & $J$& $A^{SM}$   & $J_f$  & $J_i$   &  $nlj(\rho)$& $nlj(\eta)$ & $J$& $A^{SM}$   \\
 \midrule
 $1^+$ & $0^+_1$ &  $0p_{1/2}$  & $0p_{1/2}$ & 1 & $ 0.58269$ &  $1^+$ & $2^+$   &  $0d_{3/2}$  & $0d_{5/2}$ & 3 & $ 0.00407$ \\
 $1^+$ & $0^+_2$ &  $0p_{1/2}$  & $0p_{1/2}$ & 1 & $ 0.62634$ &  $1^+$ & $2^+$   &  $0d_{3/2}$  & $1s_{1/2}$ & 1 & $ 0.06353$ \\
 $1^+$ & $0^+_1$ &  $0p_{1/2}$  & $0p_{3/2}$ & 1 & $ 0.08622$ &  $1^+$ & $2^+$   &  $0d_{3/2}$  & $1s_{1/2}$ & 2 & $-0.00157$ \\
 $1^+$ & $0^+_2$ &  $0p_{1/2}$  & $0p_{3/2}$ & 1 & $-0.04502$ &  $1^+$ & $2^+$   &  $0d_{5/2}$  & $0d_{3/2}$ & 1 & $-0.13446$ \\
 $1^+$ & $0^+_1$ &  $0p_{3/2}$  & $0p_{1/2}$ & 1 & $-0.08623$ &  $1^+$ & $2^+$   &  $0d_{5/2}$  & $0d_{3/2}$ & 2 & $ 0.04508$ \\
 $1^+$ & $0^+_2$ &  $0p_{3/2}$  & $0p_{1/2}$ & 1 & $ 0.04508$ &  $1^+$ & $2^+$   &  $0d_{5/2}$  & $0d_{3/2}$ & 3 & $-0.00406$ \\
 $1^+$ & $0^+_1$ &  $0p_{3/2}$  & $0p_{3/2}$ & 1 & $-0.09427$ &  $1^+$ & $2^+$   &  $0d_{5/2}$  & $0d_{5/2}$ & 1 & $ 0.09096$ \\
 $1^+$ & $0^+_2$ &  $0p_{3/2}$  & $0p_{3/2}$ & 1 & $ 0.02279$ &  $1^+$ & $2^+$   &  $0d_{5/2}$  & $0d_{5/2}$ & 2 & $ 0.00001$ \\
 $1^+$ & $0^+_1$ &  $0d_{3/2}$  & $0d_{3/2}$ & 1 & $ 0.03725$ &  $1^+$ & $2^+$   &  $0d_{5/2}$  & $0d_{5/2}$ & 3 & $-0.00022$ \\
 $1^+$ & $0^+_2$ &  $0d_{3/2}$  & $0d_{3/2}$ & 1 & $ 0.01689$ &  $1^+$ & $2^+$   &  $0d_{5/2}$  & $1s_{1/2}$ & 2 & $ 0.00025$ \\
 $1^+$ & $0^+_1$ &  $0d_{3/2}$  & $0d_{5/2}$ & 1 & $ 0.03982$ &  $1^+$ & $2^+$   &  $0d_{5/2}$  & $1s_{1/2}$ & 3 & $-0.00586$ \\
 $1^+$ & $0^+_2$ &  $0d_{3/2}$  & $0d_{5/2}$ & 1 & $ 0.01663$ &  $1^+$ & $2^+$   &  $1s_{1/2}$  & $0d_{3/2}$ & 1 & $-0.06353$ \\
 $1^+$ & $0^+_1$ &  $0d_{3/2}$  & $1s_{1/2}$ & 1 & $ 0.02205$ &  $1^+$ & $2^+$   &  $1s_{1/2}$  & $0d_{3/2}$ & 2 & $-0.00156$ \\
 $1^+$ & $0^+_2$ &  $0d_{3/2}$  & $1s_{1/2}$ & 1 & $ 0.01008$ &  $1^+$ & $2^+$   &  $1s_{1/2}$  & $0d_{5/2}$ & 2 & $-0.00023$ \\
 $1^+$ & $0^+_1$ &  $0d_{5/2}$  & $0d_{3/2}$ & 1 & $-0.03981$ &  $1^+$ & $2^+$   &  $1s_{1/2}$  & $0d_{5/2}$ & 3 & $-0.00587$ \\
 $1^+$ & $0^+_2$ &  $0d_{5/2}$  & $0d_{3/2}$ & 1 & $-0.01662$ &  $1^+$ & $2^+$   &  $1s_{1/2}$  & $1s_{1/2}$ & 1 & $ 0.12143$ \\
 $1^+$ & $0^+_1$ &  $0d_{5/2}$  & $0d_{5/2}$ & 1 & $ 0.00526$ &  $1^+$ & $3^-$   &  $0p_{1/2}$  & $0d_{3/2}$ & 2 & $-0.00874$ \\
 $1^+$ & $0^+_2$ &  $0d_{5/2}$  & $0d_{5/2}$ & 1 & $ 0.00847$ &  $1^+$ & $3^-$   &  $0p_{1/2}$  & $0d_{5/2}$ & 2 & $-0.02594$ \\
 $1^+$ & $0^+_1$ &  $1s_{1/2}$  & $0d_{3/2}$ & 1 & $-0.02204$ &  $1^+$ & $3^-$   &  $0p_{1/2}$  & $0d_{5/2}$ & 3 & $-0.04672$ \\
 $1^+$ & $0^+_2$ &  $1s_{1/2}$  & $0d_{3/2}$ & 1 & $-0.01008$ &  $1^+$ & $3^-$   &  $0p_{3/2}$  & $0d_{3/2}$ & 2 & $ 0.03579$ \\
 $1^+$ & $0^+_1$ &  $1s_{1/2}$  & $1s_{1/2}$ & 1 & $-0.04015$ &  $1^+$ & $3^-$   &  $0p_{3/2}$  & $0d_{3/2}$ & 3 & $-0.00462$ \\
 $1^+$ & $0^+_2$ &  $1s_{1/2}$  & $1s_{1/2}$ & 1 & $-0.03097$ &  $1^+$ & $3^-$   &  $0p_{3/2}$  & $0d_{5/2}$ & 2 & $ 0.02006$ \\
 $1^+$ & $2^+$   &  $0p_{1/2}$  & $0p_{1/2}$ & 1 & $-0.20164$ &  $1^+$ & $3^-$   &  $0p_{3/2}$  & $0d_{5/2}$ & 3 & $ 0.00563$ \\
 $1^+$ & $2^+$   &  $0p_{1/2}$  & $0p_{3/2}$ & 1 & $-0.62738$ &  $1^+$ & $3^-$   &  $0p_{3/2}$  & $0d_{5/2}$ & 4 & $-0.01783$ \\
 $1^+$ & $2^+$   &  $0p_{1/2}$  & $0p_{3/2}$ & 2 & $-0.44890$ &  $1^+$ & $3^-$   &  $0p_{3/2}$  & $1s_{1/2}$ & 2 & $-0.01519$ \\
 $1^+$ & $2^+$   &  $0p_{3/2}$  & $0p_{1/2}$ & 1 & $ 0.62744$ &  $1^+$ & $3^-$   &  $0d_{3/2}$  & $0p_{1/2}$ & 2 & $-0.00874$ \\
 $1^+$ & $2^+$   &  $0p_{3/2}$  & $0p_{1/2}$ & 2 & $-0.44895$ &  $1^+$ & $3^-$   &  $0d_{3/2}$  & $0p_{3/2}$ & 2 & $-0.03579$ \\
 $1^+$ & $2^+$   &  $0p_{3/2}$  & $0p_{3/2}$ & 1 & $-0.31358$ &  $1^+$ & $3^-$   &  $0d_{3/2}$  & $0p_{3/2}$ & 3 & $-0.00461$ \\
 $1^+$ & $2^+$   &  $0p_{3/2}$  & $0p_{3/2}$ & 3 & $-0.04167$ &  $1^+$ & $3^-$   &  $0d_{5/2}$  & $0p_{1/2}$ & 3 & $-0.04673$ \\
 $1^+$ & $2^+$   &  $0d_{3/2}$  & $0d_{3/2}$ & 1 & $ 0.00777$ &  $1^+$ & $3^-$   &  $0d_{5/2}$  & $0p_{3/2}$ & 2 & $ 0.02004$ \\
 $1^+$ & $2^+$   &  $0d_{3/2}$  & $0d_{3/2}$ & 3 & $ 0.00194$ &  $1^+$ & $3^-$   &  $0d_{5/2}$  & $0p_{3/2}$ & 3 & $-0.00560$ \\
 $1^+$ & $2^+$   &  $0d_{3/2}$  & $0d_{5/2}$ & 1 & $ 0.13445$ &  $1^+$ & $3^-$   &  $0d_{5/2}$  & $0p_{3/2}$ & 4 & $-0.01783$ \\
 $1^+$ & $2^+$   &  $0d_{3/2}$  & $0d_{5/2}$ & 2 & $ 0.04508$ &  $1^+$ & $3^-$   &  $1s_{1/2}$  & $0p_{3/2}$ & 2 & $-0.01519$ \\
\bottomrule
\end{tabular}\\
\end{table*}

Two-nucleon spectroscopic amplitudes for the target overlaps $\langle  ^{14}\textrm{N}~|~ ^{12}\textrm{C}  \rangle$ were calculated using the \texttt{KSHELL} code~\cite{shimizu2019thick} utilizing the effective nucleon-nucleon interaction \cite{yuan2012shell}. 
The results of the calculations are summarized in Table~\ref{tab:sa}.
To incorporate the calculated amplitudes into the reaction model, they were transformed into the corresponding cluster coupling scheme (e.g., see \cite{thompson1988coupled}).

 The relative‑motion wave functions were generated using a Woods–Saxon potential (Eq. \ref{eq:wsReal}) with a radius parameter of 1.25 fm and a diffuseness of 0.65 fm, while the depth was adjusted to reproduce the experimental binding energy.

For the projectile overlaps, the relative‑motion wave function of the $\alpha+d$ system was obtained by projecting the internal $s$-state of the deuteron onto the three‑body wave function as follows:
\begin{equation}
\phi^{3b}_{\alpha+d}(\textbf{y}) = \langle~ \chi_{d}(\textbf{x}) ~|~ \Psi (\textbf{x}, \textbf{y}) ~\rangle,
\label{eq:relativeMotion}
\end{equation}
where $\chi_{d}(\textbf{x})$ is the deuteron wave function, which can be numerically calculated using the RSC potential \cite{day1981three}.
The wave function $\phi^{3b}_{\alpha+d}$ exhibits an asymptotic behaviour of the form $ \exp(-kr)/r $, where $ k = \sqrt{2\mu E_b / \hbar^2} $, $\mu=4/3$ and the binding energy is $E_b = 1.474$~MeV.
The spectroscopic amplitude associated with deuteron removal from $^6$Li is derived using the previously defined wave function as follows:
\begin{equation}
A^{3b} = \int d y ~ \phi(\textbf{y})^{3b}_{\alpha+d},
\label{eq:sa}
\end{equation}
where the integration is carried out over the relative coordinate $y$ of the $\alpha+d$ system.

The angular distribution results from the CRC calculations are shown together with the experimental cross sections in Fig. \ref{fig:deuteronransferChannel} (\textit{a}).
 It can be seen that the CRC calculations reproduce the experimental data well, except at angles beyond 50$^\circ$. 
 We suggest that other reaction mechanisms, such as knock-out or breakup processes, may contribute in this region and are not explicitly included in the CRC model. 
 Nevertheless, the overall shape of the calculated cross section follows the experimental data, indicating a reasonable description of the reaction dynamics.

It is interesting to note that the contribution of the higher excited channels in the $^6$Li~+~$^{12}$C$^*$ system plays an important role in shaping the cross sections. 
This contribution becomes evident when the couplings from $^6$Li + $^{12}$C$^*$ to 
$\alpha$ + $^{14}$N are included, 
compared to the direct DWBA transition from $^6$Li + $^{12}$C$_{gs}$ to $\alpha$ + $^{14}$N (see Fig.~\ref{fig:deuteronransferChannel}). 
A particularly noticeable contribution originates from the $^6$Li + $^{12}$C$(2^{+})$ channel, which can be interpreted as a consequence of the slightly larger shell-model spectroscopic amplitude, $A^{{SM}} = 0.63$, compared to $A^{{SM}} = 0.58$ for the ground-state channel $^6$Li + $^{12}$C$_\textrm{gs}$ (see Tab. \ref{tab:sa}). 
In the case of the $^{12}$C($^6$Li,~$\alpha$)$^{14}$N reaction, the effects of strong coupled‑channels interactions are clearly manifested and are shown to be significant contributors to the formation of the cross section. Similar effects have been reported for other nuclear reactions in various studies, such as Refs. \cite{urazbekov2019clusterization, harakeh1980strong, thompson1989threshold,Lukyanov2024}.

\begin{figure}
    \centering
    \includegraphics[scale=0.6]{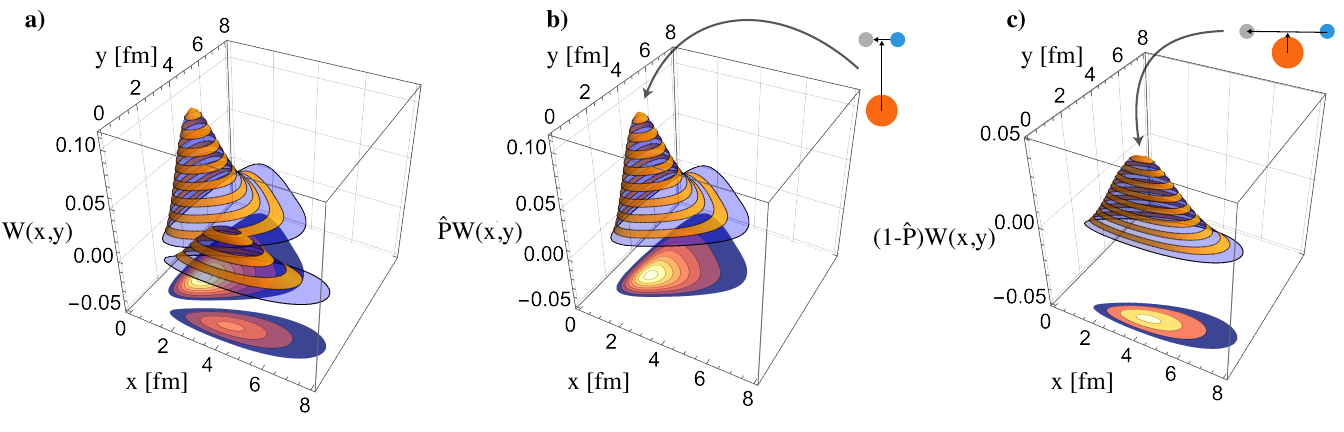}
    \caption{
3D plots of the density distribution functions $W(x,y)$ of $^6$Li within the three-body model, as implemented in the CRC calculations, shown in terms of two different spatial configurations:
a) total density distribution function;
b) the DB  configuration;
c) the CIG  configuration.
    }
    \label{fig:densityDistr}
\end{figure}

\subsection*{Correlation of the three-body structure }

It is well known that the nucleus $^6$Li exhibits two different spatial configurations, which can be clearly seen in the plot of the probability density function
\begin{equation}
W(x,y)= |\Psi(x,y)|^2 x^2 y^2.
\end{equation}
These two configurations -- the "dumb-bell," characterized by a larger distance in the $y$ coordinate than in $x$, and the "cigar-like," where the $x$ coordinate is much larger than $y$ -- are shown in Fig.~\ref{fig:densityDistr}\textit{a)}.
To isolate the  dumb-bell  (DB) configuration from the three-body wave function, we apply the operator (see e.g. in Ref. \cite{oganessian1999dynamics}):
\begin{equation}\label{eq:DB}
\hat{P}_{DB} = \frac{1}{1+\exp\left(\xi/\xi_0\right)},
\end{equation}
while the corresponding operator for the extracting the cigar-like (CIG) configuration is given by

\begin{equation}
\hat{P}_{CIG} = 1-\hat{P}_{DB} = \frac{1}{1+\exp\left(-\xi/\xi_0\right)},
\label{eq:CIG}
\end{equation}
where $\xi = (\gamma x - y)/\sqrt{1+\gamma^2}$ and its parameters were taken as $\xi_0 = 0.2$ fm and $\gamma = 0.60$.

Using the relative-motion wave functions of the $\alpha + d$ system (Eq. \ref{eq:relativeMotion}) and applying the two different projection operators (Eq. \ref{eq:DB} and \ref{eq:CIG}) to the three-body wave function $\Psi(x,y)$, we estimated the individual contributions to the reaction cross section within the CRC. 
As mentioned above, only the $S$‑wave component of the three‑body configuration was used, since it contributes most significantly.
The results of the CRC calculations with the two different spatial configurations are presented in Fig. \ref{fig:deuteronransferChannel} \textit{(b)}.
The DB configuration dominates the angular distribution across nearly the entire range, indicating that $^6$Li predominantly behaves as an $\alpha + d$ cluster structure. 
The CIG configuration contributes less overall, but it is still important to take into account. 
At forward angles, where its contribution is strongest, its weight becomes comparable to that of the DB configuration, indicating that in this region the transfer predominantly occurs in the surface-localized part. 
At larger scattering angles, the CIG contribution gradually diminishes.
Nevertheless, despite their different behaviors and oscillations in the calculated cross sections, the interference of these two configurations provides a good overall description and reproduces the experimental data reasonably well, supporting the validity of the chosen structural model of \lisix.

\begin{figure}
    \centering
        \includegraphics[scale=0.6]{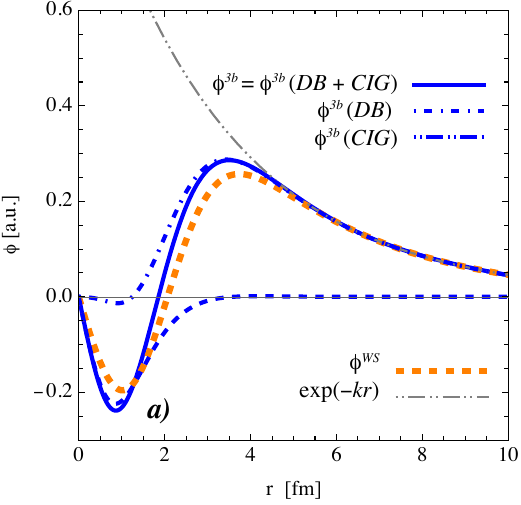}~~
         \includegraphics[scale=0.6]{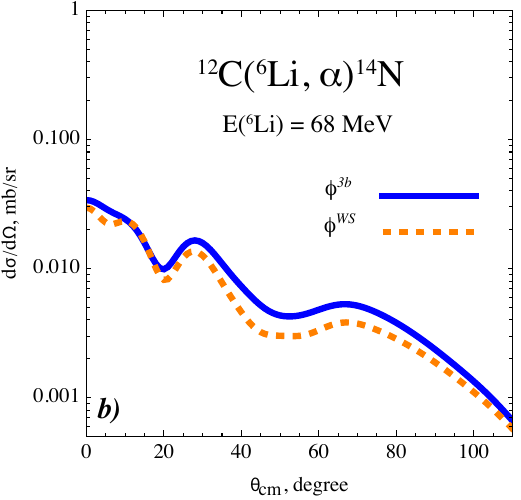}
    \caption{
Comparison of the relative‑motion wave functions obtained from the three‑body wave function $\phi^{3b}_{\alpha+d}$ with the phenomenological $\phi^{WS}_{\alpha+d}$: (a) showing the two spatial configurations (DB and CIG), both exhibiting the same asymptotic behavior $ \exp(-kr)/r $; (b) implementation of the two relative‑motion wave functions, $\phi^{3b}_{\alpha+d}$ and $\phi^{WS}_{\alpha+d}$, in the CRC calculation.
For clarity in the figure, the subscript $\alpha + d$ in $\phi^{3b}_{\alpha+d}$ and $\phi^{WS}_{\alpha+d}$ has been omitted.
   
    }
    \label{fig:relativeMotion}
\end{figure}

By splitting the relative-motion wave function $\phi^{3b}_{\alpha+d}$ into two components, it becomes possible to explain the dominance of the DB configuration. 
In addition to the greater height of the DB structure in the density distribution function (see Fig.~\ref{fig:densityDistr}~(\textit{a})), this dominance is also reflected in the relative-motion wave function of the $\alpha + d$ system, as illustrated in Fig.~\ref{fig:relativeMotion}~(\textit{a}). It can be clearly seen that the DB structure accounts for about 80\% of the integrated area $\int d y~|\phi(y)|$, while the remaining 20\% corresponds to the CIG structure.

The wave function $\phi^{3b}_{\alpha+d}$ used in the CRC calculations appears to have a bit more compact shape compared with the phenomenological wave function $\phi^{W}_{\alpha+d}$ constructed using the Woods–Saxon potential. This tendency is visible from the position of the node being slightly closer to zero and is also reflected in the rms radii, which are 2.49 fm for $\phi^{3b}_{\alpha+d}$ and 2.54 fm for $\phi^{WS}_{\alpha+d}$. 
The spectroscopic amplitude calculated with (Eq.~\ref{eq:sa}) is $A^{3b}=1.07$, whereas the phenomenological wave function gives $A^{WS}=0.97$. The three‑body spectroscopic amplitude is therefore in closer agreement with the value of $A^{SM}=1.06$ obtained within the framework of the translation-invariant shell model in Ref.\cite{rudchik1996strong} and $A^{SM}=1.09$ calculated with the \texttt{KSHELL}.

A comparison of the differential cross sections obtained with the two wave functions, $\phi^{3b}_{\alpha+d}$ and $\phi^{WS}_{\alpha+d}$, together with their corresponding spectroscopic amplitudes, is presented in Fig.~\ref{fig:relativeMotion}~(\textit{b}). Although the differences in the calculated cross sections are minor, the use of $\phi^{3b}_{\alpha+d}$ yields a slightly better agreement with the experimental data. 
This improvement is likely due to the fact that $\phi^{3b}_{\alpha+d}$ incorporates intrinsic three‑body correlations of the $^6$Li system, whereas the phenomenological $\phi^{WS}_{\alpha+d}$ does not fully include these correlations and represents the system in a more simplified manner. 
As a result, the three‑body wave function provides a more physically motivated description of the reaction dynamics.

\section*{Conclusion and outlook}
In this work, both experimental and theoretical studies of the $^{12}$C($^6$Li, $^6$Li)$^{12}$C and $^{12}$C($^6$Li, $\alpha$)$^{14}$N reactions at a laboratory energy of 68 MeV have been presented. The analysis was carried out within the OM, the CC approach, and the CRC framework, with particular attention to the cluster structure of the $^6$Li nucleus.

The entrance channel was described using the double‑folding potential based on the three‑body nuclear matter density of $^6$Li, while the exit channel was treated with the Woods–Saxon potential whose parameters were adjusted to reproduce experimental elastic‑scattering data. Spectroscopic amplitudes and relative‑motion wave functions were obtained for both phenomenological and three‑body models, and their influence on the calculated cross sections was systematically investigated.

The comparison of wave functions, including the dumb‑bell and cigar‑like spatial configurations, showed that the dumb‑bell configuration dominates over most scattering angles, while the cigar‑like configuration, though less significant overall, contributes notably at forward angles where the reaction is primarily surface‑localized. 
This highlights the intrinsic three-body correlations characteristic of $^6$Li, which are not fully captured by phenomenological models. 
The three‑body wave function was found to be more compact and to provide spectroscopic amplitudes closer to values reported in earlier studies, leading to a more consistent description of the reaction dynamics.

Coupling effects from higher excited states and inelastic channels were shown to play an important role in shaping the cross sections, especially at backward angles. The inclusion of these couplings in the CRC framework improved the agreement between calculations and experimental data, highlighting the necessity of considering both the internal structure of $^6$Li and strong coupled‑channels effects.

Overall, the results demonstrate that implementing the three‑body model of $^6$Li in CRC calculations provides a physically motivated and more accurate description of the reaction mechanisms. The interference between different spatial configurations and the explicit treatment of coupling effects are essential for understanding the underlying dynamics of light‑ion nuclear reactions.

Building on the results obtained for $^6$Li, a natural extension of this work is to investigate the $^6$He~+~$^{12}$C reaction. The nucleus $^6$He, with its Borromean structure and two-neutron halo, presents a more pronounced three-body character and offers an excellent opportunity to probe the effects of extended matter distributions and neutron correlations on reaction mechanisms. Its loosely bound nature is expected to enhance coupling to continuum states and to produce distinct signatures in transfer and breakup channels. 
Applying a similar three-body framework to $^6$He would allow a direct comparison of spatial configurations -- particularly the dineutron versus cigar-like modes -- and their influence on observables. Moreover, such studies could provide tighter constraints on reaction models involving weakly bound neutron-rich systems, offering valuable insights into both nuclear structure and dynamics beyond the stability line.

\section*{Acknowledgement}

The author (B.U.) gratefully acknowledges S.~N.~Ershov for his constructive feedback and valuable suggestions.

This research was funded by the Science Committee of the Ministry of Science and Higher Education of the Republic of Kazakhstan (Grant No. AP14870958).

\section*{References}

\bibliographystyle{iopart-num}
\bibliography{bibliography}

\end{document}